\documentclass[conference,10pt]{IEEEtran}
\IEEEoverridecommandlockouts
\usepackage{graphicx}
\usepackage{caption}
\usepackage{enumerate}
\usepackage{epsfig}
\usepackage{epstopdf}
\usepackage{cite}
\usepackage{subfigure}
\usepackage{amsmath,amssymb,amsfonts}
\usepackage{algorithmic}
\usepackage{tabularx}
\usepackage{textcomp}
\usepackage{xcolor}
\usepackage{stfloats}
\usepackage{multirow}
\usepackage{url}
\usepackage[ruled]{algorithm2e}
\usepackage{etoolbox,xstring,mfirstuc,textcase}
\usepackage{bm}
\usepackage{amsmath}
\usepackage{amsthm}
\usepackage{stackengine}
\usepackage{cuted}

\def\BibTeX{{\rm B\kern-.05em{\sc i\kern-.025em b}\kern-.08em
    T\kern-.1667em\lower.7ex\hbox{E}\kern-.125emX}}
\begin{document}

\title{Fundamental Trade-off in Wideband Stacked Intelligent Metasurface Assisted OFDMA Systems} 
	\author{Zheao Li\textsuperscript{1}, Jiancheng An\textsuperscript{1}, and Chau Yuen\textsuperscript{1 *}		\\
	\textsuperscript{1}School of Electrical and Electronic Engineering, Nanyang Technological University, 639798, Singapore.\\
		Email: zheao001@e.ntu.edu.sg,~jiancheng.an@ntu.edu.sg,~chau.yuen@ntu.edu.sg.
\vspace{-0.6 cm}
}

\maketitle
\begin{abstract}
Conventional digital beamforming for wideband multiuser orthogonal frequency-division multiplexing (OFDM) demands numerous power-hungry components, increasing hardware costs and complexity. By contrast, the stacked intelligent metasurfaces (SIM) can perform wave-based precoding at near-light speed, drastically reducing baseband overhead. However, realizing SIM-enhanced fully-analog beamforming for wideband multiuser transmissions remains challenging, as the SIM configuration has to handle interference across all subcarriers. To address this, this paper proposes a flexible subcarrier allocation strategy to fully reap the SIM-assisted fully-analog beamforming capability in an orthogonal frequency-division multiple access (OFDMA) system, where each subcarrier selectively serves one or more users to balance interference mitigation and resource utilization of SIM. We propose an iterative algorithm to jointly optimize the subcarrier assignment matrix and SIM transmission coefficients, approximating an interference-free channel for those selected subcarriers. Results show that the proposed system has low fitting errors yet allows each user to exploit more subcarriers. Further comparisons highlight a fundamental trade-off: our system achieves near-zero interference and robust data reliability without incurring the hardware burdens of digital precoding.
\end{abstract}

\vspace{-0.3 cm}
\section{Introduction}

Recent advancements in wireless communication systems demand not only higher throughput but also the capability to support diverse data streams over wide frequency bands \cite{6G}. For multiuser systems, OFDM has emerged as a fundamental technology to combat frequency-selective fading and exploit multiuser diversity \cite{OFDM}. Specifically, OFDMA extends OFDM’s advantages by partitioning subcarriers among users in the frequency domain, thereby sidestepping inter-user interference (IUI). Besides, space division multiple access (SDMA) enhances spectral efficiency through multiplexing and diversity in the spatial domain. However, SDMA typically relies on digital precoding at the base station (BS) to optimally mitigate IUI when multiple user equipments (UEs) share the same time-frequency resource block. This mandates considerable hardware cost and energy consumption in large-scale arrays and wideband transmissions, as each subcarrier requires a separate digital precoding process. 

SIMs, comprised of multiple stacked transmissive metasurface layers, enable a three-dimensional wave-based processing paradigm with significantly expanded degrees of freedom (DoF) \cite{SIM-HMIMO, SIM3}. Unlike single-layer reconfigurable intelligent surfaces (RIS), whose efficacy is largely restricted by partial reflection or refraction, the SIM leverages the successive propagation through its stacked layers to mimic fully-analog matrix operations. In essence, signals undergo ``deep neural network''-like computations at near-light speed in the electromagnetic (EM) wave domain \cite{SIM}. Compared to conventional multi-antenna architectures, an SIM-based system obviates the need for distinct digital beamforming for each subcarrier and for each transmit antenna \cite{6G2}. This can reduce the number of radio frequency (RF) chains, high-resolution digital-to-analog converters (DACs), and other power-hungry components \cite{SIM0}. Indeed, zero-forcing (ZF) beamforming across all subcarriers typically burdens digital baseband processing, with each subcarrier requiring its own precoding matrix to optimally mitigate IUI~\cite{ZF}. By contrast, an SIM can perform fully-analog beamforming with low power usage as waves traverse the stacked metasurfaces, providing a new way to coordinate interference mitigation in the wave domain.

Specifically, in \cite{BF1} and \cite{BF2}, SIM-based beamforming strategies demonstrated a reduction in energy consumption compared to conventional multiple-input multiple-output (MIMO) approaches. The authors of \cite{ICC} and \cite{ICC2} proposed SIM-aided downlink beamforming designs in multiuser systems, achieving substantial sum-rate improvements. A ZF beamfocusing approach utilizing SIM was introduced in \cite{VTC} to construct an end-to-end channel with minimal IUI. Furthermore, SIM-enhanced MIMO OFDM wideband systems were explored in \cite{Wideband-SIM, Wideband-SIM2}, where fully-analog transceiver architectures were used to enhance spatial multiplexing and mitigate deep fading, outperforming conventional wideband beamforming designs. Nonetheless, the configuration of SIM is largely fixed across multicarriers, limiting its adaptiveness in frequency-selective environments. Achieving robust, fully-analog IUI mitigation in wideband multiuser settings thus remains challenging.

To address these issues, we propose a novel framework that leverages SIM to implement fully-analog beamforming while introducing a flexible subcarrier allocation strategy for wideband OFDMA downlink transmissions. Unlike conventional approaches that rely on baseband digital precoding, our system exploits wave-based processing of SIM to perform fully-analog ZF beamforming across a broad frequency range. We dynamically allocate subcarriers among UEs in the frequency domain, thereby reducing the wave-based ZF burden on the SIM. This selective subcarrier sharing relaxes the interference-mitigation constraints in the spatial domain, improving both spectral efficiency and computational feasibility. To achieve this, we use an alternating optimization (AO) algorithm that jointly optimizes subcarrier allocation and SIM-based ZF beamforming, approximating an end-to-end IUI-free channel matrix across multicarriers. Numerical results demonstrate consistently better bit-error rate (BER) and sum rate performance of the proposed system than standard SDMA and OFDMA schemes under the same power constraints. Furthermore, our findings underscore a trade-off between subcarrier allocation and interference mitigation, pinpointing optimal performance at moderate levels of subcarrier reuse.
\vspace{-0.2 cm}
\section{The Proposed SIM-Assisted MIMO OFDMA Communication System Model}
\label{Section2}
\subsection{SIM-Assisted Transceiver Architecture for MIMO OFDMA}
\begin{figure}
	\centerline{\includegraphics[width=0.4\textwidth]{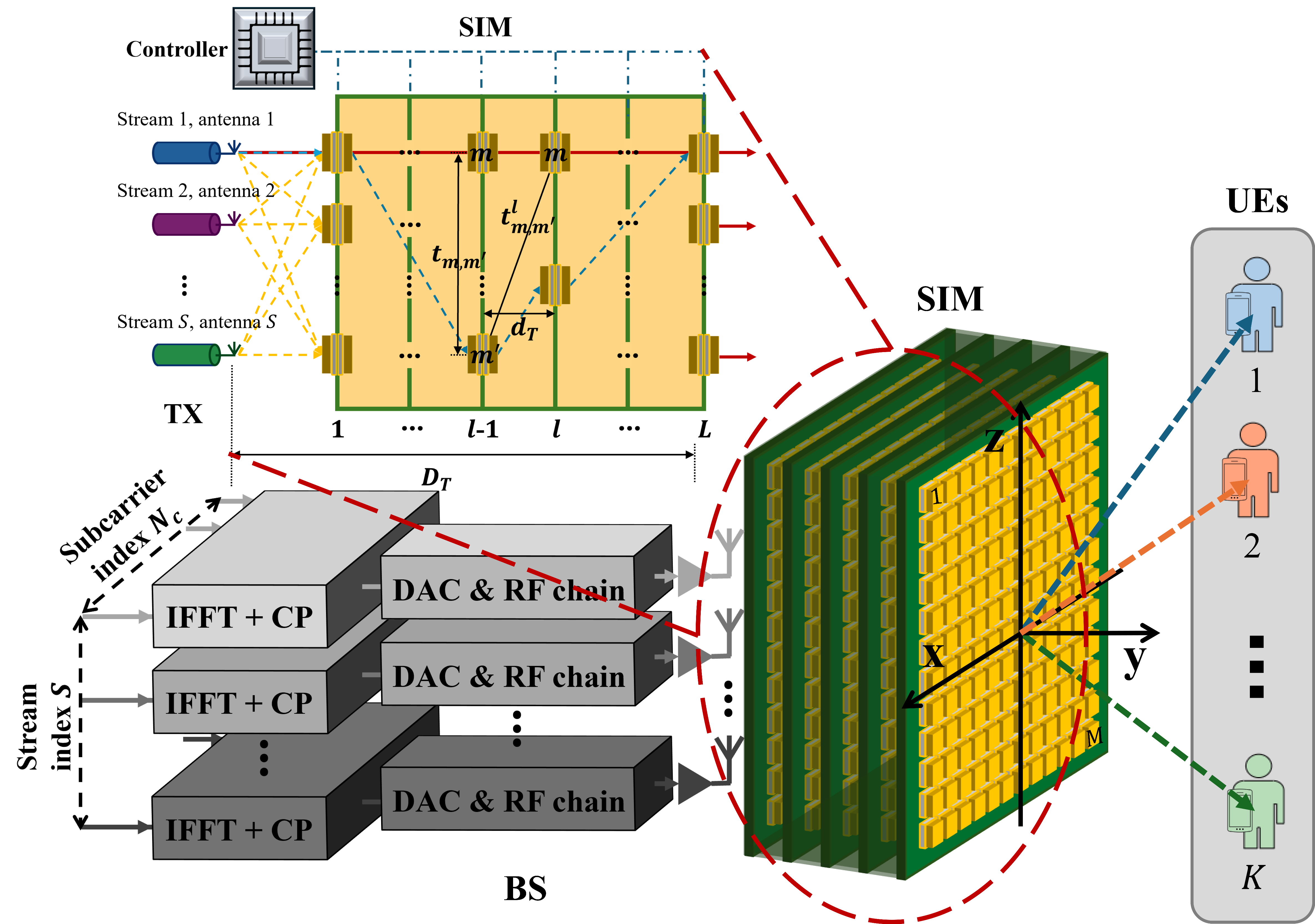}}
	\caption{\centering{SIM-assisted OFDMA downlink communication systems.}}
	\label{SIM-MU-MISO}
\vspace{-0.5 cm}
\end{figure}

Unlike conventional digital-based MIMO OFDMA architectures, which require a dedicated precoding vector for each UE, the proposed SIM-assisted MIMO OFDMA system employs an SIM positioned at the base station (BS) antenna array for performing wave-based processing to enhance signal gain and multiuser fully-analog precoding simultaneously. As shown in Fig.~\ref{SIM-MU-MISO}, the BS transmits $S$ independent data streams to $K$ single-antenna UEs. Each data stream undergoes an inverse fast Fourier transform (IFFT) operation over $N_c$ subcarriers with a cyclic prefix (CP) being appended. The processed signals are then converted to analog signals via DACs, amplified by RF chains, and transmitted through the SIM-assisted BS antenna structure. Upon reception, each single-antenna UE demodulates its respective OFDM signal. To establish multiple parallel subchannels in the physical space for interference-free transmission, the proposed SIM-assisted fully-analog system achieves the number of transmit antennas $N_{TX}= S = K$. It significantly reduces the complexity and energy consumption of baseband digital processing, eliminating the need for a large number of RF chains and enabling hardware-friendly wideband downlink communications.

Specifically, the metasurface layers of SIM dynamically interact with incident EM wavefronts in the wave domain, enabling efficient signal processing. The system comprises $L$ metasurface layers connected to intelligent controllers for performance optimization, as illustrated in Fig.~\ref{SIM-MU-MISO}. Each metasurface layer comprises $M = M_x \times M_z$ meta-atoms, where $M_x$ and $M_z$ denote the number of meta-atoms along the $x$-axis and $z$-axis, respectively. Within a single layer of the SIM, the spacing between the $m$-th and $m'$-th meta-atoms is represented as $t_{m, m'}$, while the inter-layer spacing between the $m$-th meta-atom of the $l$-th layer and the $m'$-th meta-atom of the ($l$-1)-th layer is denoted as $t_{m, m'}^l$. The overall inter-layer distance in the SIM is $d_T$, with total thicknesses denoted as $D_T$. The spacing between adjacent meta-atoms is denoted as $r_T$. 

According to the Rayleigh-Sommerfeld theory~\cite{RS}, the wideband transmission coefficients of SIM are expressed as:
\begin{align}
w_{m, m'}^l(f_i)&=\frac{S_Td_T}{{t_{m, m'}^l}^2} (\frac{1}{2\pi t_{m, m'}^l}-j\tfrac{f_i}{c})e^{j2\pi t_{m, m'}^lf_i/c},
\end{align}
where $S_T$ is the area of each meta-atom, $f_i$ denotes the frequency of the $i$-th subcarrier, and $c$ is the speed of light. 

Each meta-atom adjusts its transmission coefficient by imposing a phase shift, represented by $\phi_m^l=e^{j\theta_m^l}$ for the $m$-th meta-atom on the $l$-th layer. The phase shift matrix of the $l$-th layer is denoted by $\mathbf{\Phi}^l={\rm diag}([\phi_1^l,  \phi_2^l, \ldots, \phi_M^l]^T)\in \mathbb{C}^{M\times M}$. The cumulative effect of signal propagation through layers on subcarrier $i$ is characterized by the transmission functions:
\begin{equation}
\mathbf{P}_i=\mathbf{P}(f_i)=\mathbf{\Phi}^L\mathbf{W}^L_i \ldots \mathbf{\Phi}^2\mathbf{W}^2_i\mathbf{\Phi}^1\mathbf{W}^1_i\in \mathbb{C}^{M\times K},
\end{equation}
where $\mathbf{W}^l_i= [w_{m, m'}^l(f_i)]_{M\times M},~l=2, \ldots, L$ denotes the wideband transmission matrix between the ($l$-$1$)-th layer and the $l$-th layer of SIM and $\mathbf{W}^1_i\in \mathbb{C}^{M\times K}$ describes the transmission from transmit antennas to the first layer of SIM.

Let $\mathbf{x}_i \in \mathbb{C}^{K \times 1}$ be the transmitted signal vector containing $S=K$ streams, so the signal vector $\mathbf{y}_i \in \mathbb{C}^{K \times 1}$ received by $K$ UEs on subcarrier $i$ can be expressed as:
\begin{equation}
    \mathbf{y}_i = \mathbf{G}_i \mathbf{P}_i \mathbf{x}_i + \mathbf{n}_i,
\end{equation}
where $\mathbf{G}_i \in \mathbb{C}^{K \times M}$ is the wideband channel between the BS and UEs on subcarrier $i$. $\mathbf{x}_i = {\rm diag}(\sqrt{[p_i]_1},\ldots,\sqrt{[p_i]_K})\mathbf{s}_i$, where $[p_i]_k$ is the power allocated to the $k$-th UE on subcarrier $i$ and $\mathbf{s}_i\in \mathbb{C}^{K \times 1}$ is the data symbol vector. $\mathbf{n}_i \in \mathbb{C}^{K \times 1}$ is the additive white Gaussian noise with variance $\sigma^2_n$.

By dynamically adjusting the phase shift matrix of each metasurface layer, the SIM structure adjusts the EM response in the proposed system to perform two key functions:
\begin{enumerate}
\item \textbf{Beamforming gain}: The SIM processes the emitted wavefronts, enhancing the received power at each UE.
\item \textbf{Multiuser precoding}: The SIM spatially separates the signals destined for different UEs, mitigating IUI and optimizing signal reception at each UE.
\end{enumerate}
\vspace{-0.15 cm}
\subsection{Wideband Channel Model}
We consider a wideband multipath channel model, incorporating scatterer-based propagation and frequency-selective fading effects. The channel accounts for $P_k$ scatterers located between the SIM at BS and $k$-th UE, where the total number of paths is also $P_k$. The complex gain and delay associated with the $p$-th path are denoted as $g_p$ and $\tau_p$, respectively. For the $i$-th subcarrier with frequency $f_i$, the wideband channel vector $\mathbf{g}_k(f_i)$ between the SIM in the BS and UE $k$ is expressed as
\begin{align}
\label{channel}
\mathbf{g}_k(f_i)=  \sum_{p=0}^{P_k}g_{p}(f_i)e^{-j2\pi f_i\tau_p}{\bm{\alpha}_p(f_i)}^H \in \mathbb{C}^{1\times M}, 
\end{align}
where $\bm{\alpha}_p(f_i) \in \mathbb{C}^{M\times1}$ is the uniform planar array steering vector corresponding to the elevation angle $\vartheta_p \in [0,\pi)$ and azimuth angle $\varphi_p  \in [-\pi/2,\pi/2]$ at the SIM side.

Steering vectors $\bm{\alpha}_p(f_i)$ represents the array response of all meta-atoms in both $x$-axis and $z$-axis in the $L$-layer, which is
\begin{align}
\bm{\alpha}_p(f_i) &= \bm{\alpha}_p^{x}(f_i) \otimes \bm{\alpha}_p^{z}(f_i), \\
[\bm{\alpha}_p^{x}(f_i)] _{m_x} &\triangleq  e^{j2\pi r_{T}\sin(\vartheta_p)\sin(\varphi_p)(m_x-1)f_i/c}, \\
[\bm{\alpha}_p^{z}(f_i)] _{m_z} &\triangleq  e^{j2\pi r_{T}\cos(\vartheta_p)(m_z-1)f_i/c},
\end{align}
where $m_x = 1, 2,\ldots, M_x$ and $m_z = 1, 2,\ldots, M_z$. 

Stacking the wideband channel matrix on subcarrier $i$ from BS to all $K$ UEs yields $\mathbf{G}_i = [\mathbf{g}_1(f_i)^T,\ldots,\mathbf{g}_K(f_i)^T]^T$.

\subsection{SIM-Assisted MIMO OFDMA System Model}
To realize downlink data transmission with SDMA in a MIMO OFDM system, we directly apply the ZF precoding in the wave domain, $\mathbf{P}_i = \mathbf{G}_i^H{(\mathbf{G}_i \mathbf{G}_i^H)}^{-1}$, to mitigate IUI by
\begin{equation}
\mathbf{H}_i=\mathbf{G}_i\,\mathbf{P}_i= \mathbf{I}_{K} \in \mathbb{C}^{K \times K},
\end{equation}
where $\mathbf{H}_i\in\mathbb{C}^{K \times K}$ denotes the effective end-to-end channel on subcarrier $i$ and $\mathbf{I}_K$ the $K\times K$ identity matrix. 

\begin{figure*}[t]
    \centering
    \subfigure[Pure SDMA method, where each subcarrier is shared by all UEs.\label{fig:subfig1}]{\includegraphics[width=0.32\textwidth]{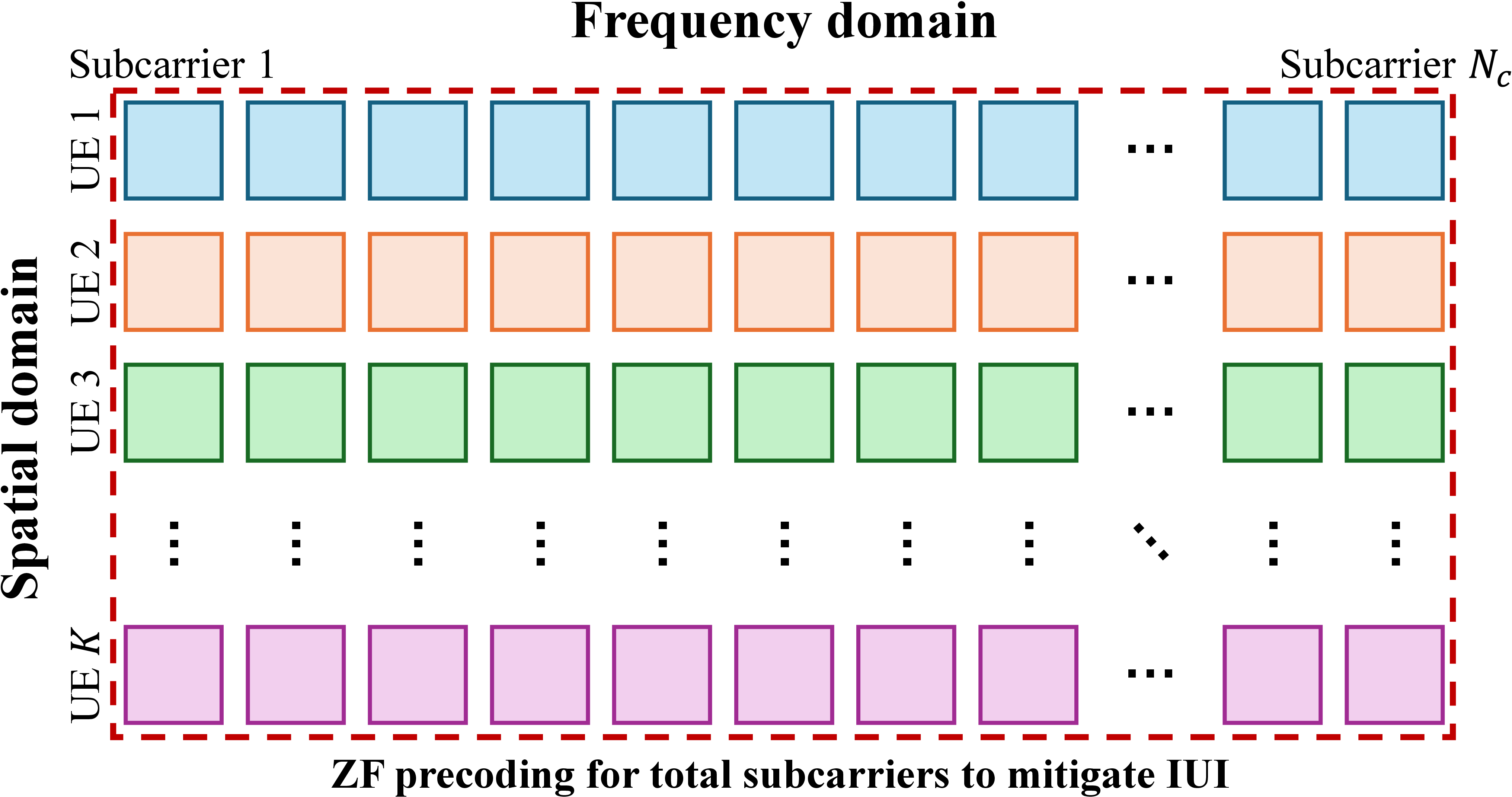}}
    \hspace{+0.1 cm}
    \subfigure[Hybrid subcarrier allocation method, grouping subcarriers into blocks with partial sharing or exclusive assignment.\label{fig:subfig2}]{\includegraphics[width=0.32\textwidth]{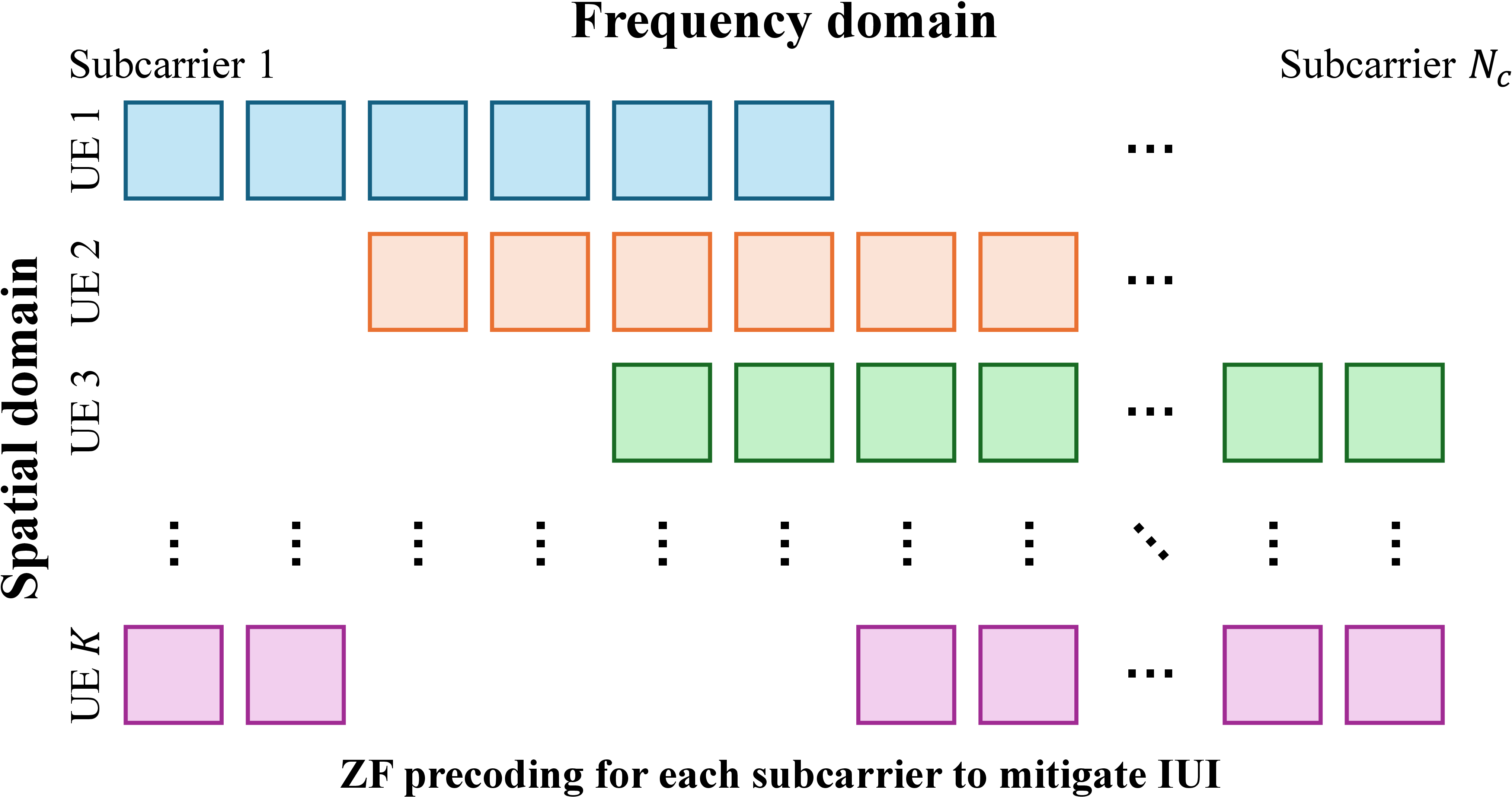}}
    \hspace{+0.1 cm}
    \subfigure[Pure OFDMA method, where each subcarrier is assigned to a single UE only.\label{fig:subfig3}]{\includegraphics[width=0.32\textwidth]{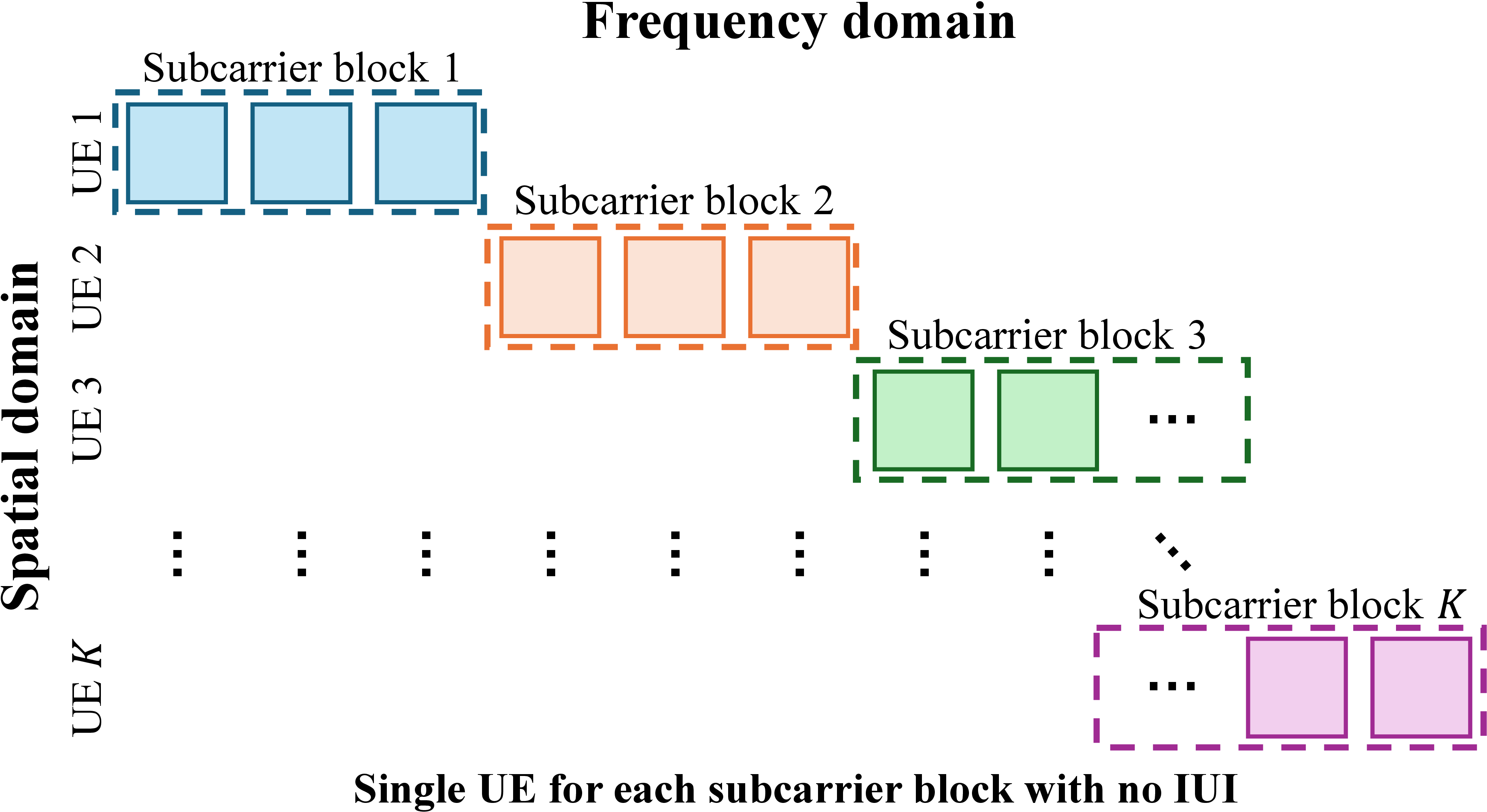}}
    \vspace{-0.2 cm}
\caption{\centering{Examples of three typical subcarrier allocation strategies.}}
    \label{fig:three-approaches}
\vspace{-0.5cm}
\end{figure*}

However, in a fully-analog wideband system assisted by SIM, high-dimensional phase shifts have to be jointly optimized across multicarriers, making a strict wave-based beamforming design extremely cumbersome and often infeasible over a very wide bandwidth \cite{Wideband-SIM2}. To alleviate this difficulty, we propose a subcarrier allocation strategy that flexibly allocates subcarriers to different UE subsets in a hybrid manner in the frequency domain and then leverages the SIM to act as a wave-based ZF precoder to mitigate IUI in the spatial domain. A binary subcarrier assignment matrix $\mathbf{Z}\in\{0,1\}^{K\times N_c}$ is introduced, where each row corresponds to a UE and each column to a subcarrier index. Specifically, $Z(k,i)=1$ means that the subcarrier $i$ is assigned to UE~$k$, whereas $Z(k,i)=0$ means that UE~$k$ does not occupy that subcarrier. Once $\mathbf{Z}$ is determined, each subcarrier knows a particular subset $\mathbf{u} = \{u_1,\dots,u_R\}^T\in \mathbb{C}^{R \times 1}$ concurrently occupying it. Assuming each UE uses $K_c$ subcarriers for transmission, the ratio of subcarrier utilization is $\rho = K_c/N_c$.

Fig.~\ref{fig:three-approaches} illustrates three typical strategies to allocate subcarriers in a multiuser system. In Fig.~\ref{fig:subfig1}, all $K$ UEs share every subcarrier, yielding the largest concurrency but also the highest IUI. A fully-analog SIM-assisted ZF precoder then has to mitigate interference across all $N_c$ subcarriers simultaneously, which can be prohibitively complex and may force constraints on the total usable bandwidth due to hardware and EM limitations of the metasurface design. Conversely, each subcarrier is exclusively assigned to a single UE in Fig.~\ref{fig:subfig3}, eliminating IUI at the cost of each UE only getting a fraction of the total subcarriers. This simplifies metasurface optimization but potentially sacrifices spectral efficiency. 

In Fig.~\ref{fig:subfig2}, by introducing the subcarrier assignment matrix $\mathbf{Z}$, we allow flexible allocation in both frequency and spatial domains. Subcarriers are partitioned into blocks, some of which may be used by one UE (like pure OFDMA) and others shared among multiple UEs (like pure SDMA). Once each subcarrier hosts only a modest subset of UEs, the SIM can then serve as a wave-based ZF precoder for that subset, thereby relaxing the optimization complexity of across-the-board interference mitigation. 
\begin{itemize}
    \item Compared to pure OFDMA, the proposed system allows higher subcarrier utilization by letting more than one UE share a subcarrier whenever beneficial, thereby boosting spectral efficiency.
  \item Compared to pure SDMA, the proposed method avoids assigning all UEs to every subcarrier, which reduces the SIM's interference mitigation burden.
\end{itemize}

The proposed method strikes flexible trade-offs by adapting subcarrier resources to channel and hardware conditions, balancing overall spectral efficiency and the IUI mitigation challenge in a fully analog SIM-assisted OFDMA system.
\section{Problem Formulation and Solution of the Joint Subcarrier Allocation and Fully-analog beamforming for MIMO OFDMA}
\label{Section3}
\subsection{Problem Formulation}
To mitigate IUI in our proposed system, we aim to ensure that the sum of the SIM-assisted end-to-end channel matrices $\mathbf{H}_i$ on each of the $N_c$ subcarriers closely approximates the desired interference-free matrix. Let $\mathbf{T}_i \in \mathbb{R}^{K \times K}$ be a diagonal selection matrix derived from $\mathbf{Z}_i$, which “activates” or “deactivates” corresponding elements of $\mathbf{G}_i$. The joint subcarrier allocation and wave-based ZF precoding optimization problem is then formulated as\noindent

\vspace{-0.2 cm}
\begin{subequations}
\label{OF_original}
\begin{align}
\label{Opt}
\mathcal{P}_1:\min_{\mathbf{Z},~\theta_m^l,~\alpha} & \mathrm{\Gamma}_{}= \sum_{i=1}^{N_c}{\bigl\|\alpha\,\mathbf{T}_i\,\mathbf{G}_i\,\mathbf{P}_i\,\mathbf{T}_i - \mathbf{T}_i \bigr\|}_F^2 \\
s.t.\quad &{\textstyle\sum_{i=1}^{N_c}} \mathbf{Z}(k,i) \;=\;  K_c,~\forall k\in \mathcal{K}, \label{st.Zs}\\
     &{\textstyle\sum_{k=1}^K}\mathbf{Z}(k,i)\;\ge\;1,~\forall i\in \mathcal{N}_c,\label{st.Zn} \\
     &\mathbf{T}_i = \mathrm{diag}\bigl(\mathbf{Z}(1,i), \mathbf{Z}(2,i),\dots, \mathbf{Z}(K,i)\bigr),\label{st.T}\\
     & \mathbf{P}_i=\mathbf{\Phi}^L\mathbf{W}^L_i \ldots \mathbf{\Phi}^2\mathbf{W}^2_i\mathbf{\Phi}^1\mathbf{W}^1_i, \label{st.Pi}\\
     &\mathbf{\Phi}^l={\rm diag}([\phi_1^l,  \phi_2^l, \ldots, \phi_M^l]^T),~\forall l\in \mathcal{L}, \label{st.pPhi} \\
     &|\phi_m^l|=|e^{j\theta_m^l}|=1,~m\in \mathcal{M},~\forall l\in \mathcal{L},  \label{st.phi}\\
     & \mathbf{Z} \in \{0,1\}^{K \times N_c}, \label{st.Z}\\
     &\alpha\in \mathbb{C},  \label{st.alpha}
\end{align}
\end{subequations}
where $\alpha$ is a scaling factor that aims to compensate for the adaptive gain in the SIM-assisted architecture for fairness. These energy-saving parts can be utilized to compensate for the energy loss caused by passing through the SIM.
\vspace{-0.2 cm}
\subsection{Solution for the Proposed Joint Optimization Algorithm}
To tackle the non-convex constraints and coupled variables in Problem $\mathcal{P}_1$, the AO algorithm is used to solve the joint optimization problems: the subcarrier allocation problem over $N_c$ subcarriers among $K$ interference channels at the BS and the phase shift optimization problem in the SIM.
\subsubsection{Optimization of the Subcarrier Allocation $\mathbf{Z}$ Given $\theta_m^l$} We aim to solve the following discrete optimization problem:
\vspace{-0.5 cm}
\begin{subequations}
\label{OF_sub}
\begin{align}
\label{Opt_sub}
\mathcal{P}_2:~~\min_{\mathbf{Z}}~~& \mathrm{\Gamma}_{}=\sum_{i=1}^{N_c}{\bigl\|\alpha\,\mathbf{T}_i\,\mathbf{H}_i\,\mathbf{T}_i - \mathbf{T}_i \bigr\|}_F^2  \\
s.t.\quad &\eqref{st.Zs},~\eqref{st.Zn},~\eqref{st.T},~\eqref{st.Z}. \label{st.all1}
\end{align}
\end{subequations}

Expanding the squared Frobenius norm in \eqref{Opt_sub} yields
\vspace{-0.2 cm}
\begin{align}
\label{eq:cost-fro}
\|\alpha\,\mathbf{T}_i\,\mathbf{H}_i\,\mathbf{T}_i - \mathbf{T}_i\|_F^2 =
{\sum_{p=1}^K} \bigl|\alpha\,\mathbf{H}_i(p,p) -1\bigr|^2\,\mathbf{Z}(p,i) + \nonumber\\
{\sum_{p\neq q}} \bigl|\alpha\,\mathbf{H}_i(p,q)\bigr|^2 \,\mathbf{Z}(p,i)\,\mathbf{Z}(q,i).
\end{align}
\vspace{-0.2 cm}

Define additional binary variables
\vspace{-0.2 cm}
\begin{align}
\mathbf{Y}(p,q,i)=\mathbf{Z}(p,i)\, \mathbf{Z}(q,i),
\;1\le p < q \le K,~\forall i\in \mathcal{N}_c. \label{st.all2}
\end{align}

The standard linearization constraints are added for each $(p,q,i)$, $p<q$ using big M method~\cite{MILP} to ensure $\mathbf{Y}(p,q,i)$ correctly models $\mathbf{Z}(p,i)\;\mathbf{Z}(q,i)$ when $\mathbf{Z}\in\{0,1\}$
\begin{align}
\mathbf{Y}(p,q,i) \;&\le\; \mathbf{Z}(p,i), \label{eq:Y1} \\
\mathbf{Y}(p,q,i) \;&\le\; \mathbf{Z}(q,i), \label{eq:Y2} \\
\mathbf{Z}(p,i) + \mathbf{Z}(q,i) &- \mathbf{Y}(p,q,i) \;\le\; 1. \label{eq:Y}
\end{align}

Specifically, $\mathbf{Z}(p,i)$ and $\mathbf{Y}(p,q,i)$ are all 0--1 decision variables, and the constraints are linear equalities and inequalities. The 0--1 quadratic objective thus forms a mixed-integer linear programme~(MILP) problem:
\begin{subequations}
\label{OF_MILP}
\begin{align}
\label{Opt_MILP}
\mathcal{P}_3:~
\min_{\mathbf{Z}}~
&\sum_{i=1}^{N_c}
\bigl[
\sum_{p=1}^K c_{p,i}\mathbf{Z}(p,i) +
\sum_{p\neq q} d_{p,q,i}\mathbf{Y}(p,q,i)
\bigr]  \\
s.t.~~
&\eqref{st.all1},~\eqref{st.all2},~\eqref{eq:Y1},~\eqref{eq:Y2},~\eqref{eq:Y},
\end{align}
\end{subequations}
where $c_{p,i} = |\alpha\,\mathbf{H}_i(p,p)-1|^2$ and $d_{p,q,i} = |\alpha\,\mathbf{H}_i(p,q)|^2$.

The resulting MILP is solved using a branch-and-bound or cutting-plane-based solver that systematically searches over binary assignments to minimize~\eqref{Opt_MILP}.

\subsubsection{Optimization of the SIM Phase Shifts $\theta_m^l$ Given $\mathbf{Z}$}
With a tentative subcarrier allocation strategy $\mathbf{Z}$, we can extract the relevant rows and columns of $\mathbf{G}_i$ and $\mathbf{P}_i$:
\begin{align}
    \mathbf{G}_i^z = \mathbf{G}_i\bigl(\mathbf{u},:\bigr) \in \mathbb{C}^{R\times M },~
    \mathbf{P}_i^z = \mathbf{P}_i\bigl(:,\mathbf{u}\bigr) \in \mathbb{C}^{M \times R}.  \label{st.P’}
\end{align}

The phase shift optimization subproblem is formulated as
\begin{subequations}
\label{OF_SIM}
\begin{align}
\label{Opt_SIM}
\mathcal{P}_4:~\min_{\theta_m^l,~\alpha}~& \mathrm{\Gamma}=\sum_{i=1}^{N_c}{\bigl\|\alpha\,\mathbf{G}_i^z\,\mathbf{P}_i^z-\mathbf{I}_{R}\bigr\|}_F^2  \\
s.t.\quad &\eqref{st.Pi},~\eqref{st.pPhi},~\eqref{st.phi},~\eqref{st.alpha},~\eqref{st.P’}. \label{st_SIM}
\end{align}
\end{subequations}

Expanding the Frobenius norm yields:
\begin{align}
\label{trace}
&{\bigl\|\alpha \mathbf{G}_i^z\mathbf{P}_i^z-\mathbf{I}_{R}\bigr\|}_F^2 =\alpha^2 {\rm Tr}(\mathbf{G}_i^z\mathbf{P}_i^z{\mathbf{P}_i^z}^H{\mathbf{G}_i^z}^H)-\nonumber\\
&\alpha {\rm Tr}(\mathbf{G}_i^z\mathbf{P}_i^z\mathbf{I}_{R})-\alpha^*{\rm Tr}(\mathbf{I}_{R}{\mathbf{P}_i^z}^H{\mathbf{G}_i^z}^H)-{\rm Tr}(\mathbf{I}_{R}).
\end{align}

Decompose $\mathbf{P}_i^z=\mathbf{P}^\mathbb{L}_i\mathbf{\Phi}^l\mathbf{P}^\mathbb{R}_i$, where $\mathbf{P}^\mathbb{L}_i=\mathbf{\Phi}^L\mathbf{W}^L...$ $\mathbf{\Phi}^{l+1}\mathbf{W}^{l+1}\in \mathbb{C}^{M\times M}$ is the product of the left-hand side of the matrix $\mathbf{P}_i^z$, and $\mathbf{P}^\mathbb{R}_i=\mathbf{W}^l\mathbf{\Phi}^{l-1}\mathbf{W}^{l-1}...\mathbf{\Phi}^{1}\mathbf{W}^{1}\bigl(:,\mathbf{u}\bigr)\in \mathbb{C}^{M\times R}$ is the product of the right-hand side of the matrix $\mathbf{P}_i^z$. By using matrix vectorization to simplify this non-trivial coupled variables, the terms of \eqref{trace} can be rewritten as
\begin{align}
\label{20}
&{\rm Tr}(\mathbf{G}_i^z\mathbf{P}_i^z{\mathbf{P}_i^z}^H{\mathbf{G}_i^z}^H)  = {\rm Tr}(\mathbf{G}_i^z\mathbf{P}^\mathbb{L}_i\mathbf{\Phi}^l\mathbf{P}^\mathbb{R}_i{\mathbf{P}^\mathbb{R}_i}^H{\mathbf{\Phi}^l}^H{\mathbf{P}^\mathbb{L}_i}^H{\mathbf{G}_i^z}^H)  \nonumber\\
& = {\rm Tr}(({\rm diag}({\bm{\phi}^l}))^H{\mathbf{P}^\mathbb{L}_i}^H{\mathbf{G}_i^z}^H\mathbf{G}_i^z\mathbf{P}^\mathbb{L}_i {\rm diag}({\bm{\phi}^l})\mathbf{P}^\mathbb{R}_i{\mathbf{P}^\mathbb{R}_i}^H) \nonumber\\
& =({\bm{\phi}^l})^H(({\mathbf{P}^\mathbb{L}_i}^H{\mathbf{G}_i^z}^H\mathbf{G}_i\mathbf{P}^\mathbb{L}_i)\odot (\mathbf{P}^\mathbb{R}_i{\mathbf{P}^\mathbb{R}_i}^H)){\bm{\phi}^l}, \\
&{\rm Tr}(\mathbf{G}_i^z\mathbf{P}_i^z) ={\rm Tr}(\mathbf{G}_i^z\mathbf{P}^\mathbb{L}_i\mathbf{\Phi}^l\mathbf{P}^\mathbb{R}_i)={\rm Tr}(\mathbf{\Phi}^l\mathbf{P}^\mathbb{R}_i\mathbf{G}_i^z\mathbf{P}^\mathbb{L}_i) \nonumber\\
&={\rm vec}({\rm diag}(\bm{\phi}^l))^H)^H {\rm vec}(\mathbf{P}^\mathbb{R}_i\mathbf{G}_i^z\mathbf{P}^\mathbb{L}_i),\\
&{\rm Tr}({\mathbf{P}_i^z}^H{\mathbf{G}_i^z}^H) = \rm Tr({\mathbf{P}^\mathbb{R}_i}^H{\mathbf{\Phi}^l}^H{\mathbf{P}^\mathbb{L}_i}^H\mathbf{G}_i^H) \nonumber\\
&=  ({\rm vec}({\rm diag}(\bm{\phi}^l)))^H{\rm vec}({\mathbf{P}^\mathbb{L}_i}^H{\mathbf{G}_i^z}^H{\mathbf{P}^\mathbb{R}_i}^H),
\end{align}
where $\bm{\phi}^l = [\phi_1^l,~\phi_2^l,~\ldots,~\phi_M^l]^T,~\forall l\in \mathcal{L}$. 

Since the quadratic terms are convex and the linear terms do not affect the convexity, the objective function \eqref{Opt_SIM} is convex. By converting unit-modulus constraints $|\phi_m^l|=|e^{j\theta_m^l}|=1$ into tractable convex forms, the unit-modulus constraint can be reformulated as ${1 \leq |\phi_m^l|}^2 \leq 1$. By introducing the non-negative slack variable $\bm{\imath}=[\imath_1, \imath_2, \ldots, \imath_{2M}]^T\in {\mathbb{R}_{\geq 0}}^{2M\times 1}$, the unit-modulus constraint \eqref{st.phi} of SIM can be relaxed by the penalty convex concave procedure (PCCP) method~\cite{PCCP}:
\vspace{-0.2 cm}
\begin{align}
{|\phi_m^l|}^2\le1+\imath_{M+m},~\forall m\in \mathcal{M}, \label{CCP_P1} \\
{|\phi_0^l|}^2-2\Re{({\phi_m^l}^*{\phi_0^l})}\le\imath_{m}-1,~\forall m\in \mathcal{M}, \label{CCP_P2}
\end{align}
where $\phi_0^l$ is the reference phase at the $l$-th layer of SIM.

The penalty factor $\rho$ is incorporated into the objective function \eqref{Opt_SIM} to minimize the values of the auxiliary variables, making the final results close to $0$. Therefore, Problem $\mathcal{P}_4$ can be transformed into Problem $\mathcal{P}_5$

\vspace{-0.5 cm}
\begin{subequations}
\label{OF_P}
\begin{align}
\mathcal{P}_5:~
\min_{{\bm{\theta}^l},~\bm{\imath}}\quad&\mathrm{\Gamma}({\bm{\theta}^l})=\mathrm{\Gamma}+\rho \sum_{i=1}^{2M}{\imath_{i}} \label{obj_P}\\
s.t.\quad &{\bm{\theta}^l} = \{\theta_m^l\}_1^M,~ \forall m\in \mathcal{M},~\forall l\in \mathcal{L} \\
&{\bm{\phi}^l} = e^{j{\bm{\theta}^l}},~\forall l\in \mathcal{L}, \label{st.ppPhi}  \\
&~\eqref{st_SIM},~\eqref{CCP_P1},~\eqref{CCP_P2} \\
& \imath_{i}\ge0,~i=1,~2, \ldots, 2M,
\end{align}
\end{subequations}
where $\mathrm{\Gamma}({\bm{\theta}^l})$ is a second-order cone programme (SOCP) prob-lem that can be solved using the CVX tool by alternating between blocks of $M$ meta-atoms with phase shifts.

To optimize the scaling factor at each iteration, the partial derivative of the objective function $\mathrm{\Gamma}$ w.r.t. $\alpha$ is calculated. Setting the derivative $\partial {\mathrm{\Gamma}}/\partial {\alpha}$ to 0 gives the unique solution:
\begin{align}
\alpha = \frac{ {\textstyle \sum_{i=1}^{N_c}} \Re({\rm tr}(\mathbf{I}_{R}\mathbf{G}_i\mathbf{P}_i))}{ {\textstyle \sum_{i=1}^{N_c}} {||\mathbf{G}_i\mathbf{P}_i||}_F^2}.
\end{align}
\vspace{-0.6cm}
\subsection{Complexity Analysis}
\vspace{-0.1cm}
The computational complexity of the proposed joint algorithm is derived from subcarrier allocation and SIM phase shift optimizations. As a 0--1 quadratic optimization, allocating these UEs into subcarriers is NP-complete. Constraints \eqref{st.Zs} and \eqref{st.Zn} further couple the row-wise variables. Thus, the complexity of branch-and-bound per iteration is $O(N_cK+N_c K^2)$. In addition, the major complexity using the PCCP method is from the SOCP problems~\eqref{OF_P}. There are $2M$ SOC constraints with dimension one per PCCP iteration, where all $M$ meta-atoms in $L$ layer of SIM have to be updated iteratively. Thus, by omitting lower-order terms, the computational complexity per iteration of solving SOCP problems is $O(N_c L M [2M+(2M)^{3.5}])$. Therefore, the total complexity of the proposed joint algorithm is $O(I_{AO} N_c (L M^{4.5}+K^2))$, where $I_{AO}$ is the number of AO iterations.
\vspace{-0.2cm}
\section{Results and Analysis}
\vspace{-0.2cm}
\label{Section5}
The system operates at a center frequency $f_0 = 28$ GHz over a bandwidth $B = 40$ MHz and the number of subcarriers is set as $N_c = 16$. As shown in Fig.~\ref{SIM-MU-MISO}, the BS has $S$ antennas in a uniform linear array, equipped with $L$-layer SIM having $M$ meta-atoms per layer. The height of BS is $10$ m, and the vertical distance between BS and $K$ UEs is $250$ m, with each UE evenly spaced $30$ m apart. The thickness of SIM is set as $D_T=0.05$ m, and the spacing between the layers is set as $d_T = D_T/L$. The parameters of SIM are configured with $S = K=4$, $M = 100$, $L= 7$, $r_{T}= c/(2 f_0)$, and $S_{T}= c^2/(2f_0)^2$. $100$ scatterers are randomly generated to model channel characteristics. The antenna gains of the BS and UEs are set to $3$ dBi and $0$~dBi, respectively. The noise power spectral density is $-112$~dBm/Hz. The BS power allocation vector $[p_i]_k$ is optimized based on the iterative water-filling algorithm~\cite{Wideband-SIM2}. Each initialization runs for iterations $I_{AO}=50$, with dynamic adjustment based on convergence. To ensure reliability, the Montecarlo method is utilized to average the outcomes over 100 simulation runs. 

\begin{figure}[t]
    \centering
    \subfigure[Fitting NMSE versus the number of subcarriers used per UE $K_c$.\label{NMSE:subfig1}]{\includegraphics[width=0.23\textwidth]{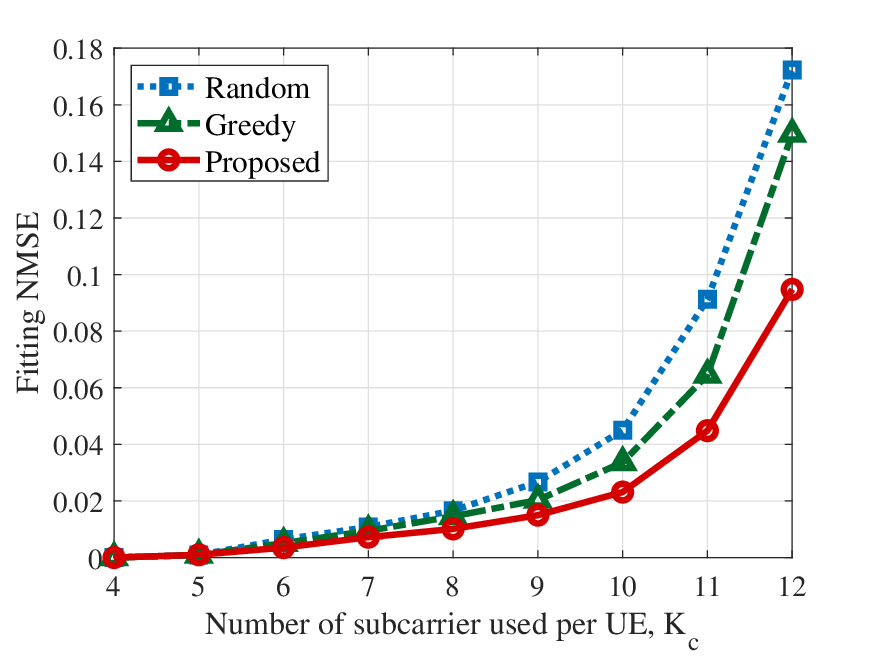}}
    \hspace{+0.15 cm}
    \subfigure[End-to-end channel visualization when $\rho=0.625$.\label{NMSE:subfig2}]{\includegraphics[width=0.18\textwidth]{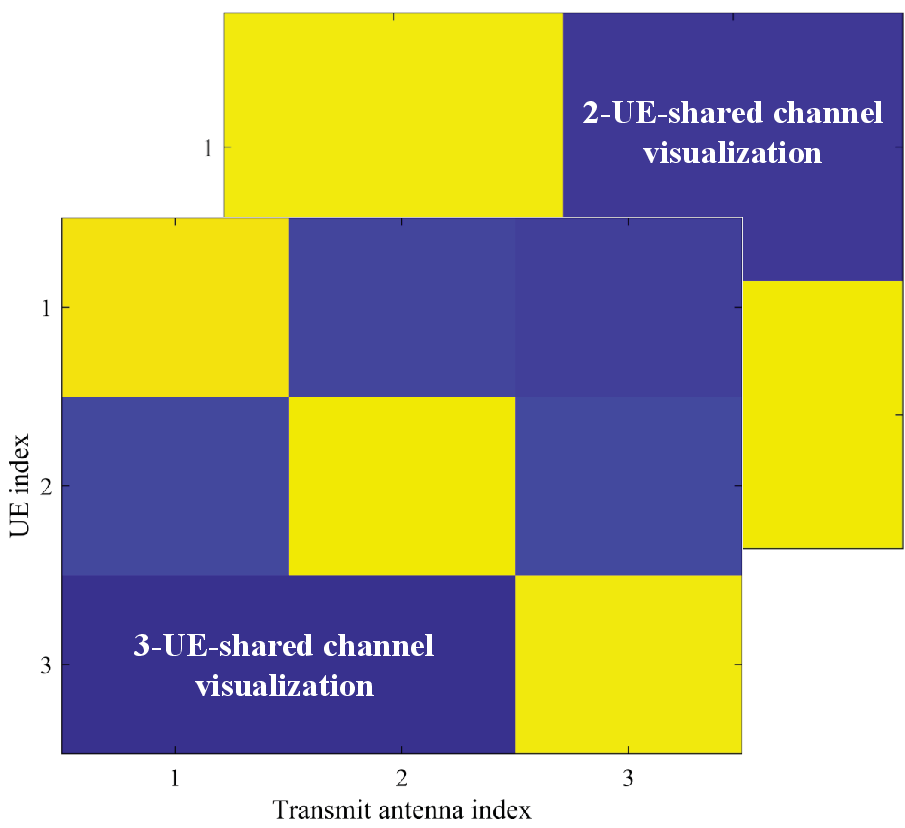}}
    \vspace{-0.2 cm}
\caption{Results of the proposed SIM-assisted joint optimization.}
	\label{NMSE} 
\vspace{-0.7cm}
\end{figure}

In Fig.~\ref{NMSE}, the proposed joint optimization scheme is compared with two baseline strategies for generating the subcarrier assignment matrix $\mathbf{Z}$. In the random approach, subcarriers are assigned to each user purely at random according to a target ratio. In the greedy approach, for each subcarrier, we first rank the candidate UEs by their peak channel magnitudes, then tentatively add the highest-gain pair to $\mathbf{Z}$ if it does not overly degrade the partial wave-domain fitting. This subcarrier-by-subcarrier assignment continues until a preset ratio of subcarrier usage per UE is reached. After each baseline fixes $\mathbf{Z}$, the SIM phase shift matrices are optimized. As shown in Fig.~\ref{NMSE:subfig1}, plotting the fitting normalized mean square error (NMSE) versus the number of subcarriers used per UE, the proposed design consistently achieves the lowest NMSE across all tested $K_c$. At $K_c =10$, each UE uses 10 out of 16 subcarriers with $\rho = 0.625$, which is the maximum ratio of subcarrier utilization to preserve near-zero interference among shared subcarriers. The heatmap in Fig. \ref{NMSE:subfig2} demonstrates how the end-to-end channel approximates an identity matrix for each partially shared subcarrier block, confirming that the SIM can effectively align user streams as needed. Consequently, the proposed joint optimization of $\mathbf{Z}$ and $\mathbf{\Phi}$ not only achieves higher subcarrier usage, but also preserves near-zero IUI. These results underscore that our joint strategy provides a more favorable balance between bandwidth utilization and interference mitigation, leading to significantly lower fitting errors than either random or greedy assignment alone.

\begin{figure}
	\centerline{\includegraphics[width=0.3\textwidth]{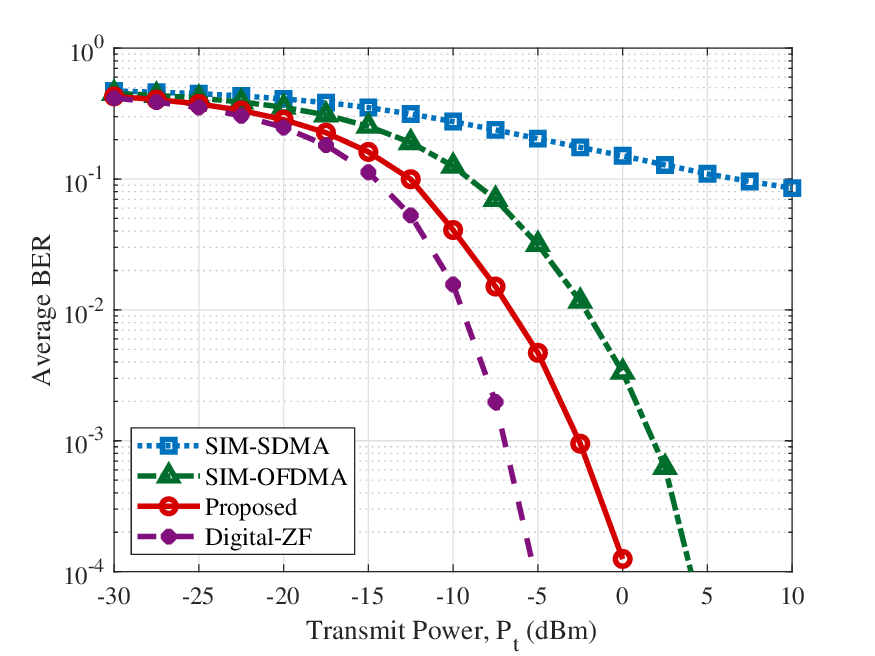}}
\vspace{-0.1cm}
	\caption{{Average BER comparison of SIM-SDMA, SIM-OFDMA, digital-ZF, and the proposed SIM-assisted system.}}
	\label{BER} 
\vspace{-0.5cm}
\end{figure}

Fig. \ref{BER} presents the BER performance of the proposed SIM-assisted system with a ratio of subcarrier utilization $\rho = 0.625$, compared to three benchmark schemes: SIM-assisted SDMA, SIM-assisted OFDMA, and digital-ZF precoding. The system employs binary phase shift keying (BPSK) modulation with four independent data streams, each corresponding to a distinct UE. To ensure a fair comparison, an SNR offset is applied to the pure SIM-OFDMA scheme to compensate for its lower spatial multiplexing gain and transmission rate, as each UE occupies fewer subcarriers. Under this normalization, OFDMA eliminates IUI but suffers from reduced effective SNR, resulting in inferior BER compared to the proposed method in the medium-to-high transmit power regime. Meanwhile, SIM-SDMA exhibits the worst performance due to severe IUI arising from full subcarrier reuse, as the fully-analog SIM-based ZF precoding cannot be fully optimized across a wide bandwidth. In contrast, the proposed scheme integrates a flexible subcarrier-user assignment matrix with wave-based SIM precoding, effectively suppressing IUI while maintaining high subcarrier utilization. This joint design yields superior BER, when the transmit power exceeds $-15$ dBm. Although digital-ZF achieves slightly better BER at high power levels, the proposed fully-analog solution offers a more favorable trade-off between hardware complexity and performance.

\begin{figure}
	\centerline{\includegraphics[width=0.3\textwidth]{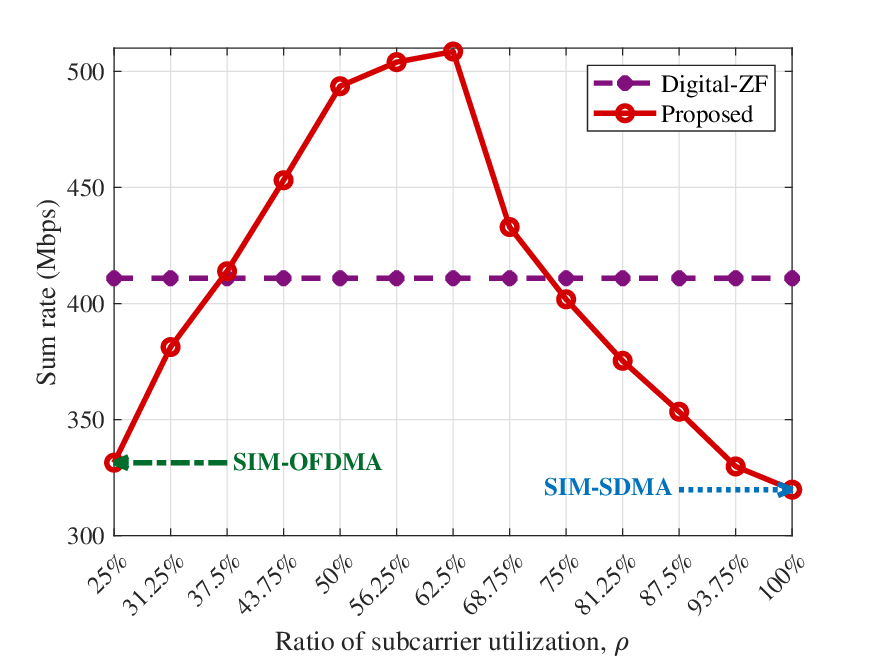}}
\vspace{-0.1cm}
	\caption{{Sum rate comparison between the proposed SIM-assisted system and the conventional digital-ZF baseline.}}
	\label{sumrate} 
\vspace{-0.7cm}
\end{figure}

Fig.~\ref{sumrate} then examines the sum rate under different $\rho$ under $10$ dBm power. As $\rho$ increases, more UEs share each subcarrier, improving spectral efficiency by leveraging spatial‐frequency multiplexing. However, once $\rho$ becomes excessively large, the SIM’s ability to suppress IUI weakens due to physical limitations in wave-based precoding, resulting in a degraded sum rate. This reveals a fundamental trade-off: low $\rho$ underutilizes spatial resources like SIM-OFDMA, while high $\rho$ suffers from excessive interference like SIM-SDMA. The proposed method achieves its maximum sum rate around $\rho=62.5\%$, where subcarrier reuse and interference control are optimally balanced. Notably, in the range $K_c \in [6, 12)$, the proposed system outperforms the digital-ZF baseline. This performance gain stems from the large-aperture SIM architecture, which not only aligns wavefronts for beamforming but enhances power concentration in desired directions. These results demonstrate that, when properly designed, fully-analog SIM-assisted precoding can outperform that with digital-ZF in terms of both system performance and hardware efficiency.
\section{Conclusions}
\label{Section6}
This paper has presented a novel fully-analog approach for wideband MIMO OFDMA by jointly optimizing subcarrier allocation and SIM coefficient design. By dynamically determining the subcarrier allocation and simultaneously optimizing the SIM’s phase shifts, the system has effectively approximated near-zero interference at significantly lower complexity and power usage than conventional digital precoding. The results have shown that the partial subcarrier‐sharing strategy, combined with an iterative SIM phase‐optimization, achieves lower fitting errors with more subcarriers per UE than either greedy or random‐based allocation methods. Moreover, compared to standard SDMA/OFDMA and digital-ZF precoding baselines, the proposed system has provided robust BER and sum rate performance under identical power constraints without requiring digital precoding. Overall, this system has offered~an efficient trade-off between spectral efficiency, IUI mitigation, and hardware simplicity for next-generation wideband wireless communication networks.

\vspace{-0.2cm}
	\section*{Acknowledgment}
	\small {This work was supported by A*STAR (Agency for Science, Technology and Research) Singapore, under Grant No. M22L1b0110, and the Ministry of Education (MOE), Singapore, under its MOE Tier 2 Award MOE-T2EP50124-0032.}


\vspace{-0.2cm}
\begin{thebibliography}{}
\vspace{-0.1cm}
\bibitem{6G} 
C. -X. Wang \emph{et al.}, “On the road to 6G: Visions, requirements, key technologies and testbeds,” \emph{IEEE Commun. Surveys Tuts.}, vol. 25, no. 2, pp. 905--974, 2nd Quart. 2023.

\bibitem{OFDM} 
G. L. Stuber, J. R. Barry, S. W. McLaughlin, Ye Li, M. A. Ingram, and T. G. Pratt, “Broadband MIMO-OFDM wireless communications," \emph{Proc. of the IEEE}, vol. 92, no. 2, pp. 271--294, Feb. 2004.

\bibitem{SIM-HMIMO} 
J. An \emph{et al.}, “Stacked intelligent metasurfaces for efficient holographic MIMO communications in 6G," \emph{IEEE J. Sel. Areas Commun.}, vol. 41, no.~8, pp. 2380--2396, Aug. 2023.

\bibitem{SIM3} 
J. An \emph{et al.}, “Emerging technologies in intelligent metasurfaces: Shaping the future of wireless communications,” \emph{IEEE Trans. Antennas Propag.}, doi: 10.1109/TAP.2025.3571069.

\bibitem{SIM} 
C. Liu \emph{et al.}, “A programmable diffractive deep neural network based on a digital-coding metasurface array,” \emph{Nature Electron.}, vol. 5, no. 2, pp.~113--122, Feb. 2022.

\bibitem{6G2} 
J. An, C. Yuen, L. Dai, M. Di Renzo, M. Debbah, and L. Hanzo, “Near-field communications: Research advances, potential, and challenges,” \emph{IEEE Wireless Commun.}, vol. 31, no. 3, pp. 100--107, June 2024.

\bibitem{SIM0} 
J. An \emph{et al.}, “Stacked intelligent metasurface-aided MIMO transceiver design," \emph{IEEE Wireless Commun.}, vol. 31, no. 4, pp.~123--131, Aug. 2024.

\bibitem{ZF} 
E. Björnson, M. Bengtsson, and B. Ottersten, “Optimal multiuser transmit beamforming: A difficult problem with a simple solution structure,” \emph{IEEE Signal Process. Mag.}, vol. 31, no. 4, pp. 142--148, July 2014.

\bibitem{BF1}
N. U. Hassan \emph{et al.}, “Efficient beamforming and radiation pattern control using stacked intelligent metasurfaces," \emph{IEEE Open J. Commun. Society}, vol. 5, pp. 599--611, Jan. 2024.

\bibitem{BF2}
A. Papazafeiropoulos, P. Kourtessis, S. Chatzinotas, D. I. Kaklamani, and I. S. Venieris, “Near-field beamforming for stacked intelligent metasurfaces-assisted MIMO networks," \emph{IEEE Wireless Commun. Letters}, vol. 13, no. 11, pp. 3035--3039, Nov. 2024.

\bibitem{ICC} 
J. An, M. Di Renzo, M. Debbah, and C. Yuen, “Stacked intelligent metasurfaces for multiuser beamforming in the wave domain,” in \emph{Proc. IEEE ICC}, Rome, Italy, May 2023, pp. 1--6.

\bibitem{ICC2} 
J. An \emph{et al.}, "Stacked intelligent metasurfaces for multiuser downlink beamforming in the wave domain," \emph{IEEE Trans. Wireless Commun.}, vol. 24, no. 7, pp. 5525--5538, July 2025.

\bibitem{VTC} 
X. Jia \emph{et al.}, "Stacked intelligent metasurface enabled near-field multiuser beamfocusing in the wave domain," in \emph{Proc. IEEE VTC-Spring}, Singapore, Singapore, June 2024, pp. 1--5.

\bibitem{Wideband-SIM}
Z. Li, J. An, and C. Yuen, “Stacked intelligent metasurfaces for fully-analog wideband beamforming design,” in \emph{Proc. IEEE VTS APWCS}, Singapore, Singapore, Aug. 2024, pp. 1--5.

\bibitem{Wideband-SIM2}
Z. Li, J. An, and C. Yuen, “Stacked intelligent metasurfaces-enhanced MIMO OFDM wideband communication systems,” \emph{arXiv:2503.00368}, Mar. 2025.

\bibitem{RS} 
X. Lin \emph{et al.}, “All-optical machine learning using diffractive deep neural networks,” \emph{Sci.}, vol. 361, no. 6406, pp. 1004--1008, July 2018.

\bibitem{MILP} 
H. P. Williams, \emph{Model building in mathematical programming}. John Wiley \& Sons, 2013.

\bibitem{PCCP} 
G. Zhou, C. Pan, H. Ren, K. Wang, and A. Nallanathan, “A framework of robust transmission design for IRS-aided MISO communications with imperfect cascaded channels,” \emph{IEEE Trans. Signal Process.}, vol. 68, pp.~5092--5106, Aug. 2020.
\end{thebibliography}
\end{document}